\documentclass[prb,twocolumn,floatfix,showpacs,preprintnumbers,superscriptaddress,amsmath,amssymb]{revtex4}
\usepackage[dvips]{graphics}
\usepackage{psfrag}
\usepackage{graphicx}
\def\iv{{$I-V$ curve$\;$}}

\begin{document}

\title{Shot noise in tunneling transport through molecules and quantum dots} 

\author {Axel Thielmann} 
\affiliation{Forschungszentrum Karlsruhe, Institut f\"ur Nanotechnologie,
76021 Karlsruhe, Germany}

\author {Matthias H. Hettler} 
\affiliation{Forschungszentrum Karlsruhe, Institut f\"ur Nanotechnologie,
76021 Karlsruhe, Germany}

\author {J\"urgen K\"onig} 
\affiliation{Institut f\"ur Theoretische Festk\"orperphysik,
Universit\"at Karlsruhe, 76128 Karlsruhe, Germany}

\author {Gerd Sch\"on} 
\affiliation{Institut f\"ur Theoretische Festk\"orperphysik,
Universit\"at Karlsruhe, 76128 Karlsruhe, Germany}

\date{\today}

\begin{abstract}
We consider electrical transport through single molecules coupled to metal 
electrodes via tunneling barriers.
Approximating the molecule by the Anderson impurity model as the simplest 
model which includes Coulomb charging effects, we extend the ``orthodox'' 
theory to expand current and shot noise systematically order by order in 
the tunnel couplings.
In particular, we show that a combined measurement of current and shot noise
reveals detailed information of the system even in the weak-coupling 
limit, such as the ratio of the tunnel-coupling strengths of the molecule to 
the left and right electrode, and the presence of the Coulomb charging energy.
Our analysis holds for single-level quantum dots as well.
\end{abstract}

\maketitle

\section{Introduction}
 
Single-molecule electronics provides an exciting possibility for further 
miniaturization of electronic devices.
Nonlinear electrical transport through single molecules deposited between 
mechanically controlled break junctions has been measured in
several experiments.
\cite{reed-etal,kergueris-etal,reichert-etal,platin-complex} 
Furthermore, recently observed diode behavior\cite{metzger,vilan}
memory effects and negative differential conductance\cite{chen-reed-cp}
in devices which make use of molecular films have attracted much 
interest.

A crucial problem for understanding and designing molecular electronic devices
is the lack of control of the metal-molecule interfaces.
In molecular-film experiments, deposition of the covering electrode is 
very difficult and can easily destroy the film or induce short cuts.
Once the molecules are deposited, no direct study and manipulation of the 
contact is possible.
Even in a break junction experiment one has very limited influence 
on how a molecule bonds to the two electrodes.
Information on the coupling strengths $\Gamma_{\rm L}, \Gamma_{\rm R}$
between the molecule and electrodes has to  rely on transport measurements. 
From the measurement of the current $I$, however, only a combination of 
$\Gamma_{\rm L}$ and $\Gamma_{\rm R}$ can be deduced, whereas the
asymmetry ratio $\Gamma_{\rm L}/ \Gamma_{\rm R}$ remains undetermined.

In addition, it is impossible to
judge from the \iv alone, whether charging effects such as Coulomb
blockade are relevant to the transport. This is because any observed set
of current steps expected for an \iv displaying Coulomb blockade can
be mimicked by transport through a set of noninteracting levels.
As so far there is no quantitative theory  for describing
hybrid metal-molecule functions, it remains a matter of opinion,
whether such a fit is reasonable for a given experiment.

Further information can be gained from the shot noise $S$, allowing, in 
principle, to determine the ratio of the couplings. 
Furthermore, the combination of current and shot-noise measurement  
provides information about the presence and size of Coulomb charging energy.
For transport through a noninteracting 
level\cite{Khlus,Lesovik,blanter}  
the so-called Fano factor $F= S/2eI$ is known to be 
$(\Gamma_{\rm L}^2 + \Gamma_{\rm R}^2)/(\Gamma_{\rm L} +\Gamma_{\rm R})^2$. 
However, transport through weakly coupled molecules can not be modeled by 
noninteracting molecular levels, since Coulomb blockade effects become 
important.\cite{hsw,hettler_prl,cocomplex}

Some results have been obtained
before for shot noise through interacting 
systems.\cite{hershfield,nazarov,loss,bagrets,haug} 
However, the focus in these works was on transport through impurity
states or quantum dots, where some questions pertinent 
to molecules are irrelevant.

In the present paper, we reformulate the ``orthodox'' theory of shot
noise\cite{blanter} in a way that allows a  
systematic perturbation expansion in the coupling strength.
We study in detail the shot noise for a molecule accommodating one level,
described by the single-level Anderson model, in the lowest-order perturbation 
theory.
We perform a complete classification of all transport regimes, and evaluate 
the corresponding Fano factors, recovering known expressions for the cases
previously studied in the literature.
The Fano factor turns out to depend in a nonmonotonic way on the asymmetry 
ratio $\Gamma_{\rm L}/\Gamma_{\rm R}$.
The $\Gamma_{\rm L}/\Gamma_{\rm R}$ dependence of the Fano factor can be
used to identify the different transport regimes.
Our results hold for transport through single-level quantum dots as well.
In contrast to molecular devices, the contacts to leads in the 
semiconductor quantum dots are usually well controllable.

\section{The model}
The simplest model for electron transport through a molecule with Coulomb
interaction is the Anderson-impurity model described by the Hamiltonian
$\hat H = \hat H_{\rm L} + \hat H_{\rm R} + \hat H_{\rm Mol} + 
\hat H_{\rm T,L} + \hat H_{\rm T,R}$ with 
\begin{eqnarray}
  && \hat H_{r} = \sum_{k \sigma}\epsilon_{k \sigma r} a_{k \sigma r}^{\dag}
  a_{k \sigma r}, \\
  && \hat H_{\rm Mol} = \sum_{\sigma}\epsilon_{\sigma} c_{\sigma}^{\dag} 
  c_{\sigma} + U n_{\uparrow}n_{\downarrow},\\
  && \hat H_{{\rm T},r}= \sum_{k \sigma}(t_r a_{k \sigma r}^{\dag} c_{\sigma} 
  + h.c.)
\end{eqnarray}
and $r = {\rm L,R}$.
Here, $\hat H_{\rm L}$ and $\hat H_{\rm R}$ model the left and right 
electrodes with noninteracting electrons, $\hat H_{\rm Mol}$ describes the 
molecule with one relevant (spin-dependent) molecular level of energy 
$\epsilon_{\sigma}$, and the Coulomb interaction $U$ on the molecule
($n_{\uparrow}$ and $n_{\downarrow}$ are the number operators for electrons 
with corresponding spin).
Tunneling between leads and molecule level is modeled by 
$\hat H_{\rm T,L}$ and $\hat H_{\rm T,R}$.
The coupling strength is characterized by the intrinsic linewidth 
$\Gamma_r = 2\pi \rho_e |t_r|^2$, where $\rho_e$ is the (constant) 
density of states of the leads.
Furthermore, we define the total linewidth $\Gamma = \Gamma_{\rm L} + 
\Gamma_{\rm R}$.
The Fermi operators $a_{k \sigma r}^{\dag} (a_{k \sigma r})$ and
$c_{\sigma}^{\dag} (c_{\sigma})$ create (annihilate) electrons in the 
electrodes and the molecule.

We are interested in transport through the molecule, in particular in the
current $I$ and the (zero-frequency) current noise $S$.
They are related to the current operator 
$\hat I = (\hat I_{\rm R} - \hat I_{\rm L})/2$, with $\hat{I}_r = -i(e/\hbar) 
\sum_{k \sigma} \left( t_r a_{k \sigma r}^{\dag} c_{\sigma} - h.c.\right)$ 
being the current operator for electrons tunneling into lead $r$, by 
$I = \langle \hat{I} \rangle$, and 
\begin{equation}
S = \int_{-\infty}^{\infty} dt \langle \delta \hat{I}(t) \delta \hat I(0) 
+ \delta \hat{I}(0) \delta \hat I(t) \rangle ,
\end{equation}
where 
$\delta \hat I(t)=\hat I(t)-\langle \hat I \rangle$.

\section{Diagrammatic Technique}

A diagrammatic technique to perform a systematic perturbation expansion of 
the current through localized levels has been developed in 
Ref.~\onlinecite{diagrams}.
Here, we expand this technique to address the perturbation expansion of the 
current noise as well.
After this general formulation, we specify the results for lowest order
in $\Gamma$ and discuss implications for transport through molecules in this
regime.

All transport properties are governed by the nonequilibrium time evolution of
the density matrix.
The leads are described as reservoirs of noninteracting electrons, which
remain in thermal equilibrium.
A finite transport voltage enters the difference of the electrochemical 
potentials $\mu_r$ for the left and right leads, $r={\rm L,R}$.
These degrees of freedom can be integrated out making use of Wick's theorem,
such that the Fermion operators are contracted in pairs.
As a result, we end up with a reduced density matrix for the molecule's degrees of
freedom only, labeled by $\chi$.
In our case, four molecule states are possible: the level is empty, $\chi=0$, 
singly occupied with either spin up or down, $\chi = \uparrow$ or $\downarrow$,
or doubly occupied $\chi=d$. 
The time evolution of the reduced density matrix, described by the propagator
$\Pi_{\chi' \chi}(t',t)$ for the propagation from state $\chi$ at time $t$ to
state $\chi'$ at time $t'$, can be visualized by diagrams.

\begin{figure}[h]
\centerline{\includegraphics[width=8cm]{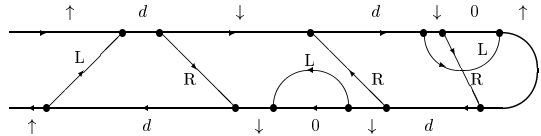}}
\caption{
  An example for the time evolution of the reduced density matrix. 
  The upper and lower lines represent the forward and backward
  time propagation along the Keldysh contour, 
  respectively. Tunneling lines correspond to the reservoirs $\rm L,R$
  connecting pairs of vertices. The resulting changes between the four
  molecular states are indicated.}
\label{fig:keldysh}
\end{figure}

An example is shown in Fig.~\ref{fig:keldysh}.
Vertices representing tunneling, $\hat H_{\rm T}$, are connected in pairs by 
tunneling lines.
The full propagation is expressed as a sequence of irreducible blocks 
$W_{\chi' \chi}(t',t)$ containing one or more tunneling lines.
\begin{figure}[h]
\centerline{\includegraphics[width=7cm]{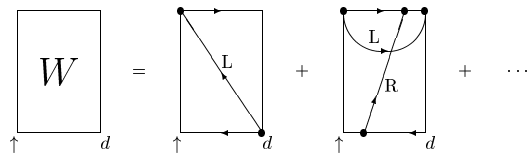}}
\caption{An example for an irreducible block $W$.}
\label{fig:W-rates}
\end{figure}
They are associated with transitions from state $\chi$ at time $t$ to state
$\chi'$ at time $t'$ (see Fig.~\ref{fig:W-rates}).
This leads to the Dyson equation for the propagator
\begin{equation}
  {\bf \Pi}(0,t) = {\bf 1} + \int_t^{0} dt_2 \int_{t}^{t_2} dt_1 \,
  {\bf W}(t_2,t_1) {\bf \Pi}(t_1,t) ,
\label{eq:dyson}
\end{equation}
where the boldface indicates matrix notation related to the molecular state
labels.
To calculate transport properties in the stationary limit, we start at time 
$t \rightarrow -\infty$, at which the system has a given diagonal but 
otherwise arbitrary distribution of probabilities $p_{\chi}^{\rm init}$ to be 
in state $\chi$, comprised in the vector ${\bf p}^{\rm init}$.
The stationary probability distribution 
${\bf p}^{\rm st} = {\bf \Pi}(0,-\infty) {\bf p}^{\rm init}$, however,
does not depend on ${\bf p}^{\rm init}$, i.e.,
$\Pi_{\chi'\chi}(0,-\infty) = p_{\chi'}^{\rm st}$, independent of $\chi$.
To determine ${\bf p}^{\rm st}$ we use the Dyson equation Eq.~(\ref{eq:dyson})
in the limit $t\rightarrow -\infty$, multiply ${\bf p}^{\rm init}$ from
the right, and define ${\bf W} = \int_{-\infty}^0 dt\,\hbar {\bf W} (0,t)$
to get ${\bf p}^{\rm st} - {\bf p}^{\rm init} = {1\over \eta} {\bf W} 
{\bf p}^{\rm st}$, where $\eta \rightarrow 0^+$ comes from a convergence
factor $\exp(-\eta |t_2|/\hbar)$ in the time integral.
This leads to ${\bf W} {\bf p}^{\rm st} = \bf 0$
together with the normalization ${\bf e}^T {\bf p}^{\rm st} = 1$, where
the vector $\bf e$ is given by $e_\chi = 1$ for all $\chi$.
The matrix $\bf W$ has zero determinant and hence
cannot be inverted (this can be seen from the sum rule
${\bf e}^T {\bf W} = {\bf 0}$).
To write down a single matrix equation which determines 
${\bf p}^{\rm  st}$ we make use of the normalization of probabilities and 
introduce the matrix $\bf \tilde W$, which is identical to $\bf W$ but with 
one (arbitrarily chosen) row $\chi_0$ being replaced with 
$(\Gamma,...,\Gamma)$.
The stationary probabilities are the solution of the linear equation
\begin{equation}
  ( {\bf \tilde W} {\bf p}^{\rm st} )_\chi 
  = \Gamma \delta_{\chi, \chi_0}  \, .
\end{equation}
This can be solved for ${\bf p}^{\rm st}$ by inverting $\bf \tilde W$.

For a well-defined perturbation expansion in powers $k$ of the coupling 
strength $\Gamma$ we write ${\bf W} = \sum_{k=1}^{\infty} {\bf W}^{(k)}$, 
${\bf \tilde W} = \sum_{k=1}^{\infty} {\bf \tilde W}^{(k)}$, and 
${\bf p}^{\rm st} = \sum_{k=0}^{\infty} {\bf p}^{{\rm st}(k)}$.
The order in $\Gamma$ corresponds to the number of tunnel lines
contained in the irreducible blocks.
As a consequence, $\bf W$ and ${\bf \tilde W}$ start with first order in 
$\Gamma$, whereas ${\bf p}^{\rm st}$ starts in zeroth order.
The zeroth-order stationary probabilities are
\begin{equation}
  p^{{\rm st}(0)}_\chi= (\tilde W^{(1)})^{-1}_{\chi \chi_0} \Gamma \, 
\label{eq:pst_0}
\end{equation}
and higher-order corrections are obtained iteratively by
\begin{equation}
  {\bf p}^{{\rm st}(k)}= -\left( {\bf \tilde W}^{(1)} \right)^{-1} 
  \sum_{m=0}^{k-1} {\bf \tilde W}^{(k-m+1)} {\bf p}^{{\rm st}(m)} \, 
\label{eq:pst}
\end{equation}
for $k = 1,2,\ldots$.
The diagrammatic rules for the calculation of the blocks ${\bf W}^{(k)}$ 
will be summarized at the end of this section.

The formulation of the perturbation expansion of the stationary probabilities 
is the first important step for a perturbation expansion of the current.
The diagrammatic representation of contributions to the current involve 
a block ${\bf W}^I$, in which one (internal) vertex is replaced by an
external vertex for the current operator $\hat I$ divided by $e/\hbar$.
The current can also be expanded order by order in the coupling
strength $\Gamma$,
\begin{equation}
  I^{(k)} = {e\over 2\hbar} {\bf e}^T \sum_{m=0}^{k-1} {\bf W}^{I {(k-m)}}
  {\bf p}^{{\rm st}(m)}
\label{eq:Ik}
\end{equation}
for $k=1,2,\ldots$, where the factor $1/2$ corrects for double counting of
the external vertex being on the upper and lower branch of the Keldysh contour.

Now, we turn to the evaluation of the current noise
$S = 2 \int_{-\infty}^0 dt \, [ \langle \hat I(t)\hat I(0)+
\hat I(0)\hat I(t) \rangle - 2 \langle \hat I \rangle^2 ]$, which 
involves expectation values of two current operators at different times.
These two current operators can either appear in the same irreducible block,
denoted by ${\bf W}^{II}$, or in two different blocks.
In total, the current noise is expressed as
\begin{equation}
  S = {e^2\over \hbar} {\bf e}^T \left( 
    {\bf W}^{II} {\bf p}^{\rm st} + {\bf W}^{I} 
    {\bf P} {\bf W}^{I} {\bf p}^{\rm st} \right) \, ,
\label{noise_ptilde}
\end{equation}
(where the factor $2$ in the above definition has canceled against
the correction factor $1/2$ for double counting) with
\begin{equation}
  {\bf P} = \int_{-\infty}^0 dt \,\frac{1}{\hbar} \left[ {\bf \Pi} (0,t) - 
    {\bf \Pi}(0,-\infty) \right] \, .
\label{eq:P}
\end{equation}
The constant part ${\bf \Pi}(0,-\infty)$ in the integrand is associated with
the $\langle \hat I \rangle^2$ contribution to the noise. 
Furthermore, it guarantees that the integral is convergent and that $\bf P$ 
exists.

We also can formulate a perturbation expansion of $\bf P$ in orders of 
$\Gamma$.
To do so, we first relate $\bf P$ to the blocks $\bf W$ and the stationary
probabilities ${\bf p}^{\rm st}$ by
\begin{equation}
  {\bf W P} = {\bf p}^{\rm st} \otimes {\bf e}^T - {\bf 1} \, ,
\label{eq:wp}
\end{equation}
which follows from the Dyson equation and ${\bf W p}^{\rm st} = {\bf 0}$.
As discussed above, $\bf W$ is not invertible, i.e., in addition to 
Eq.~(\ref{eq:wp}) we need some extra condition to determine $\bf P$.
This extra condition is ${\bf e}^T {\bf P} = {\bf 0}$, which follows from 
the definition, Eq.~(\ref{eq:P}), together with the Dyson equation,
Eq.~(\ref{eq:dyson}), and ${\bf e}^T {\bf W} = {\bf 0}$.
Introducing the matrix $\bf Q$ with elements given by
\begin{equation}
Q_{\chi' \chi} = (p_{\chi'}^{\rm st}-\delta_{\chi', \chi}) 
(1-\delta_{\chi',\chi_0})\, ,
\end{equation}
Eq.~(\ref{eq:wp}) can be recast as
${\bf \tilde W} {\bf P} = {\bf Q}$,
which allows the determination of $\bf P$ by inversion of the matrix 
$\bf \tilde W$ introduced above.
The solution for $\bf P$ can be expanded in orders of $\Gamma$.
We observe that, since the expansion of $\bf Q$ starts in zeroth order in 
$\Gamma$, $\bf P$ starts in order $\Gamma^{-1}$,
\begin{equation}
  {\bf P}^{(-1)} = ( {\bf \tilde W}^{(1)} ) ^{-1} 
  {\bf Q}^{(0)} \, .
\label{eq:Pm1}
\end{equation}
This reflects the fact that between the external current vertices the system
evolves in time, as described by ${\bf \Pi}(0,t)$.
The time scale at which the dot relaxes and correlations decay is set by the 
inverse of the tunnel rates, i.e., by $\hbar/\Gamma$.
As a consequence, the time integral in Eq.~(\ref{eq:P}) is effectively cut
at this time scale, i.e., the dominant contribution to ${\bf P}$ is of order 
$\Gamma^{-1}$.
Furthermore, it is crucial that ${\bf P}$ starts in order $\Gamma^{-1}$, 
as otherwise the second term in Eq.~(\ref{noise_ptilde}) would not 
contribute to noise to first order in $\Gamma$
(cf. Ref.~\onlinecite{korotkov}), leading to a wrong result even
for non-interacting systems ($U=0$), where exact expressions are available
for comparison.\cite{hershfield}

Corrections of higher order, $k = 0,1,...$, can be computed from
\begin{equation}
  {\bf P}^{(k)} = \left( {\bf \tilde W}^{(1)} \right)^{-1} 
  \left[{\bf Q}^{(k+1)} -
  \sum_{m = -1}^{k-1} {\bf \tilde W}^{(k-m+1)} {\bf P}^{(m)} \right] .
\label{eq:Pk}
\end{equation}
Eventually, we have all ingredients at hand to expand the noise in orders of
$\Gamma$, and we find
\begin{widetext}
\begin{equation}
  S^{(k)} = {e^2\over \hbar} {\bf e}^T \sum_{m=0}^{k-1} \left( 
    {\bf W}^{II (k-m)} {\bf p}^{{\rm st}(m)}
    + \sum_{m'=1}^{k-m}\,\,\,\sum_{m''=-1}^{k-m-m'-1}
    {\bf W}^{I (k-m-m'-m'')} {\bf P}^{(m'')} {\bf W}^{I (m')} 
    {\bf p}^{{\rm st}(m)} 
  \right) \, 
\label{eq:Sk}
\end{equation}
\end{widetext}
for $k=1,2,\ldots$.
In lowest order this result simplifies to
$S^{(1)} = (e^2/\hbar) \, {\bf e}^T \left( {\bf W}^{II (1)} 
+ {\bf W}^{I (1)} {\bf P}^{(-1)} {\bf W}^{I (1)}\right) {\bf p}^{{\rm st}(0)}$.

In  summary of the technical part, we have derived a scheme to perform a
systematic perturbation expansion of the current and the current noise.
First, one calculates the irreducible blocks $\bf W$, ${\bf W}^I$, and
${\bf W}^{II}$.
Second, the stationary probabilities ${\bf p}^{\rm st}$ are obtained from
Eqs.~(\ref{eq:pst_0}) and (\ref{eq:pst}) and $\bf P$ follows from
Eqs.~(\ref{eq:Pm1}) and (\ref{eq:Pk}).
Eventually, current and current noise are given by Eqs.~(\ref{eq:Ik}) and
(\ref{eq:Sk}).

The irreducible blocks $\bf W$ are calculated according to the rules specified
in Ref.~\onlinecite{diagrams} [where the irreducible blocks 
were denoted by $\bf \Sigma$ and included a factor $i$ in the definition, 
${\bf \Sigma} = i {\bf W}$]. 
We complete this section with a summary of these diagrammatic rules, given as follows:

(1) For a given order $k$ draw all topologically different diagrams with
$2k$ vertices connected by $k$ tunneling lines (an example of a first- and 
second-order diagram is shown in Fig.~\ref{fig:W-rates}).
Assign the energies $\epsilon_{\chi}$ to the propagators and
energies $\omega_l$ ($l=1,...,k$) to the tunneling lines.

(2) For each of the ($2k-1$) segments enclosed by two adjacent vertices
there is a resolvent $1/(\Delta E_j+i0^+)$, with $j=1,...,2k-1$, where 
$\Delta E_j$ is the difference of the left-going minus the right-going energies.

(3) The contribution of a tunneling line of reservoir $r$ is 
$(1/2\pi) \Gamma_r f(\omega_l-\mu_r)$ if the line is going backward with 
respect to the closed time path and $(1/2\pi) \Gamma_r [1-f(\omega_l-\mu_r)]$
if it is going forward.
Here, $f(x)$ is the Fermi function.

(4) There is an overall prefactor $(-i)(-1)^c$, where $c$ is the 
total number of vertices on the backward propagator plus the number of 
crossings of tunneling lines plus the number of vertices connecting the 
state $d$ with $\uparrow$.

(5) Integrate over the energies $\omega_l$ of the tunneling lines and sum over 
all reservoir and spin indices. 

The blocks ${\bf W}^I$ and ${\bf W}^{II}$ are determined in a similar way.
The only difference to $\bf W$ is that in ${\bf W}^I$ (${\bf W}^{II}$) one 
(two) internal vertices are replaced by external ones representing
$\hat I \hbar /e$.
This amounts to multiplying an overall prefactor, which arises due to the
definition of the current operator and since the number of internal vertices 
on the backward propagator may have changed.
We get a factor $+1/2$ for each external vertex on the upper
(lower) branch of the Keldysh contour which describes tunneling of an electron 
into the right (left) or out of the left (right) lead, and $-1/2$ in
the other four cases.
Finally, we have to sum up all the factors for each possibility to replace
one (two) internal vertices by external ones.

\begin{figure}[t]

\centerline{\includegraphics[width=7.5cm,angle=270,]{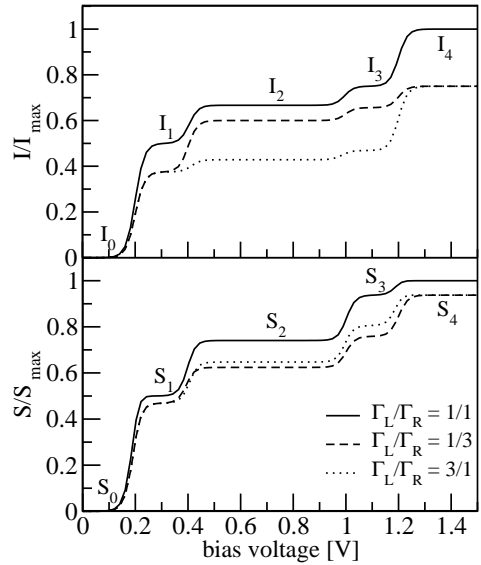}}
\caption{Current $I$ and shot noise $S$ vs. voltage for $T=100 {\rm K}$, 
  $\epsilon_{\downarrow}=0.1 {\rm eV}$, $\epsilon_{\uparrow}=0.2 {\rm eV}$,
  $U=0.4 {\rm eV}$ and $\Gamma_{\rm L} / \Gamma_{\rm R}=$ 1/1 (solid line), 1/3
  (dashed line), and 3/1 (dotted line). 
  The height of the plateaus labeled by $i=0,..,4$ are given in 
  Table~\ref{tab:table1}.
  The curves are normalized to $I_{\rm max}=(e/\hbar)\Gamma/2$ and 
  $S_{\rm max}=(e^2/\hbar)\Gamma/2$, respectively.}
\label{fig:i-s}
\end{figure}

\section{Results}

In the following we discuss shot noise in the lowest- (first-) 
order perturbation theory in $\Gamma$. 
We consider transport through a single level and allow for a finite
spin splitting. Such a spin splitting could be realized 
due to the Zeeman effect of an external magnetic
field (for semiconductor quantum dots) or an intrinsic exchange field 
due to atoms with magnetic moments (for molecules). 
The molecule can acquire four possible states: the level being unoccupied,
occupied with either spin $\uparrow$ or $\downarrow$, or doubly occupied.
The molecular level is characterized by level energies $\epsilon_\downarrow$ 
and $\epsilon_\uparrow$, the Coulomb repulsion $U$, and the coupling strengths
$\Gamma_{\rm L}$ and $\Gamma_{\rm R}$ to the electrodes.
In addition, the electron distributions in the metal electrodes are governed by
electrochemical potentials $\mu_{\rm L}$ and $\mu_{\rm R}$ and temperature $T$.
We choose a symmetric bias voltage, such that $\mu_{\rm L}=eV/2$ and 
$\mu_{\rm R}=-eV/2$. In this simple model without real spatial extent,
the voltage is dropped entirely at the  electrode-molecule tunnel junctions, 
meaning that the energies of the molecular states are independent of 
the applied voltage even if the couplings are not symmetric. For
asymmetric coupling this might not be entirely realistic. However, 
we do not wish to include a heuristic parameter to describe the
possibly unequal drop of the bias. In a comparison to an actual
experiment this could be easily amended. We also do not include an
explicit gate voltage, as its effects are straightforward to
anticipate.

In Fig.~\ref{fig:i-s} we plot the current and the current noise for a special 
choice of system parameters as a function of transport voltage.
For this figure we consider a set of energy parameters 
($\epsilon_{\downarrow}=0.1 {\rm eV}$,$\epsilon_{\uparrow}=0.2 {\rm eV}$,
$U=0.4 {\rm eV}$) that is comparable to energies encountered in
small molecules. A rescaling of energies by a
factor 1/100 would be necessary to obtain energies of the order
achievable in semiconductor quantum dots and by external magnetic
fields. If the same scaling is applied to the couplings and the
temperature, the figure would remain unchanged. This is a special feature
of first-order transport.

Electron transport becomes possible when charge excitations on the
molecule become energetically allowed. Generally, at low bias, 
transport is exponentially suppressed, unless a degeneracy of states
with different net charge is present (this could be tailored by
application of a gate voltage, which we do not consider here).
Each time when one of the four excitation energies $\epsilon_{\downarrow}$, 
$\epsilon_{\uparrow}$, $\epsilon_{\downarrow}+U$, or $\epsilon_{\uparrow}+U$ 
enters the energy window defined by the electrochemical potentials of the
electrodes, a transport channel opens. 
This gives rise to plateaus, separated by thermally broadened steps.
Due to the symmetric application of the bias, the steps occur at 
voltages of twice the corresponding excitation energy.
The plateau heights depend on the coupling parameters $\Gamma_{\rm L}$ and 
$\Gamma_{\rm R}$ only, i.e., they are independent of $U$ and $T$.
The analytic expressions for current, noise, and Fano factor of these plateaus,
which are labeled by $i=0,...,4$, are given in Table~\ref{tab:table1}.

The curves in Fig.~\ref{fig:i-s} are normalized to 
$I_{\rm max}=(e/\hbar)\Gamma/2$ and $S_{\rm max}=(e^2/\hbar)\Gamma/2$, 
respectively, which is reached in the large-bias
limit for symmetric coupling $\Gamma_{\rm L} = \Gamma_{\rm R}$.
For asymmetric coupling, the plateaus are reduced in height.
In Fig.~\ref{fig:i-s}, we show the results for 
$3\Gamma_{\rm L}=\Gamma_{\rm R}$ (dashed lines) and 
$\Gamma_{\rm L}=3\Gamma_{\rm R}$ (dotted lines) together with the case of
symmetric coupling.
The symmetry of our setup implies that all plateau heights are invariant under
simultaneous exchange of $\Gamma_{\rm L}$ with $\Gamma_{\rm R}$ and
$\mu_{\rm L}$ with $\mu_{\rm R}$.
However, the plateau height can change if only $\Gamma_{\rm L}$ and 
$\Gamma_{\rm R}$ are exchanged, or if only the bias voltage is
reversed, as shown in our example for the two plateaus labeled by $2$ and 
$3$. This opens the possibility to access the asymmetry 
$\Gamma_{\rm L}/\Gamma_{\rm R}$ experimentally by reversing the bias voltage 
and comparing the plateau heights.
\begin{widetext}

\begin{table}[h]
\centerline{\includegraphics[width=16cm]{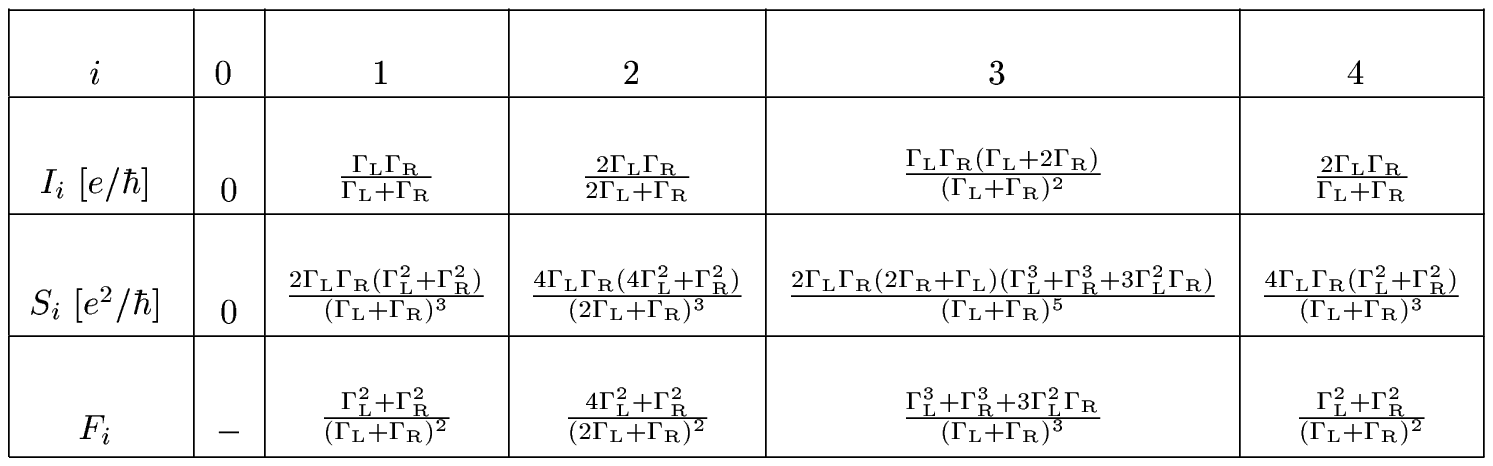}}
\caption{Current, shot noise and Fano factor for the different plateaus
  in the current-voltage characteristic shown in Fig.~\ref{fig:i-s}. 
The plateau values depend only 
on the coupling parameters $\Gamma_{\rm L,R}$.}
\label{tab:table1}
\end{table}
\end{widetext}
Figure~\ref{fig:i-s} shows the result for one special choice of 
energy parameters and corresponding excitation energies.
Nevertheless, Table~\ref{tab:table1} is complete in the sense that it 
contains all 
possible plateau values  for any configuration of the excitation
energies relative to the electrochemical potentials of the electrodes.
The classification of the configurations and the algorithm to find the
corresponding analytic expressions in Table~\ref{tab:table1} is given in 
Table~\ref{tab:table-plateaus}.
Without loss of generality we restrict ourselves to $U \geq 0$ and 
$\epsilon_{\uparrow} \geq \epsilon_{\downarrow}$.
The different configurations are classified by specifying which excitation 
energies lie within the energy window defined by the chemical potentials 
$\mu_{\rm L},\mu_{\rm R}$.
We find 13 different possibilities, as listed in 
Table~\ref{tab:table-plateaus}.
For each case, the index $i$ indicates the column where the corresponding
analytic expressions for current, noise and Fano factor can be found in 
Table~\ref{tab:table1}.
The indices $2^*$ and $3^*$ refer to columns $2$ and $3$ but with 
$\Gamma_{\rm L}$ and $\Gamma_{\rm R}$ being exchanged.

\begin{table}[h]
\centerline{\includegraphics[width=5cm]{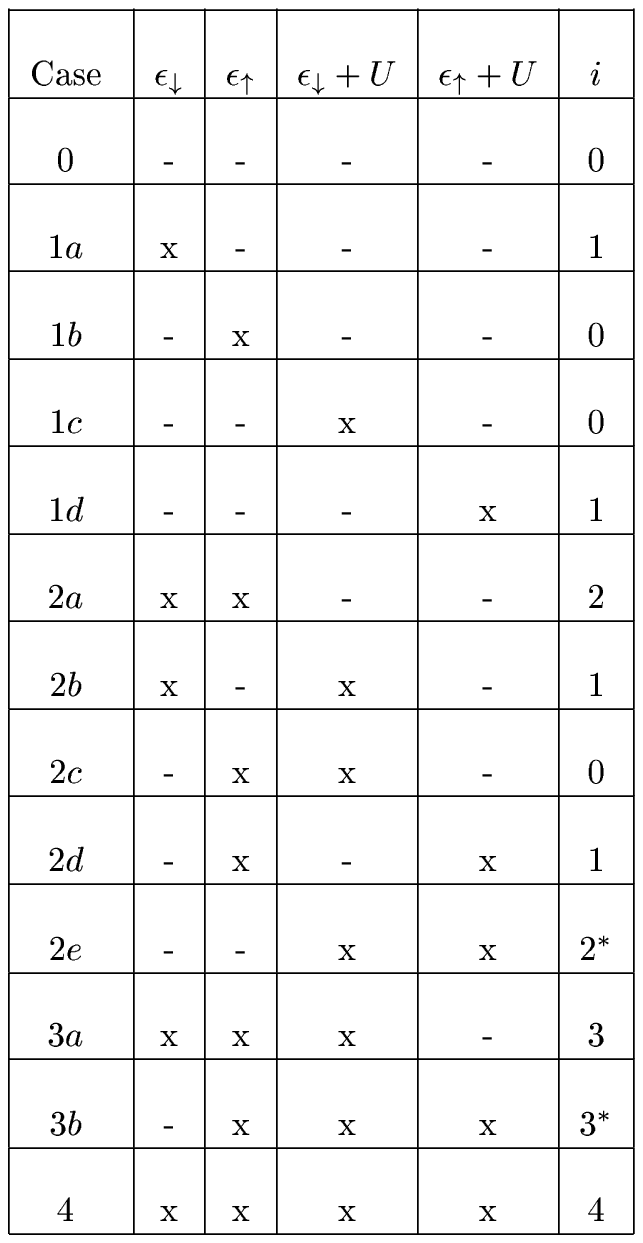}}
\caption{
  Classification of all possible configurations that are possible for 
  $U \geq 0$ and $\epsilon_{\uparrow} \geq \epsilon_{\downarrow}$.
  A cross (x) or minus ($-$) indicates that the corresponding excitation 
  energy lies within or outside the energy window defined by the 
  electrochemical potentials $\mu_{\rm L}$ and $\mu_{\rm R}$, respectively.
  For each configuration, the analytic expression for current, noise and Fano 
  factor can be found in Table~\ref{tab:table1} in column $i$.
  The indices $2^*$ and $3^*$ refer to column $2$ and $3$ with 
  $\Gamma_{\rm L}$ and $\Gamma_{\rm R}$ being exchanged.}
\label{tab:table-plateaus}
\end{table}

\begin{figure}[h]
\centerline{\includegraphics[width=6cm,angle=0]{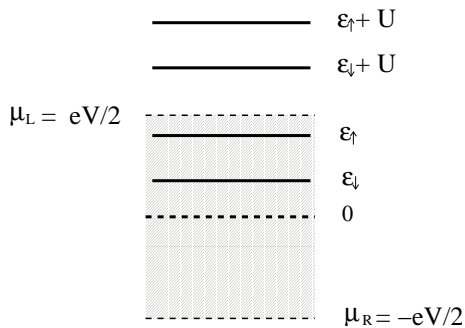}}
\caption{A sketch of the configuration $2a$ listed in 
  Table~\ref{tab:table-plateaus}.
  The excitation energies $\epsilon_{\downarrow}$ and $\epsilon_{\uparrow}$
  lie in the energy window defined by the electrochemical potentials 
  $\mu_{\rm L}$ and $\mu_{\rm R}$, and the energies $\epsilon_{\downarrow}+U$ 
  and $\epsilon_{\uparrow}+U$ lie outside.}
\label{fig:example}
\end{figure}

In order to illustrate the use of the table we sketch the situation $2a$ in 
Fig.~\ref{fig:example}, realized in Fig.~\ref{fig:i-s} in
the region between $0.4 {\rm V}$ and $1 {\rm V}$. In this situation
transport through both spin states is present as the molecule gets
charged/uncharged in the sequential tunneling events. Double occupancy 
is still out of reach, since the excitation energies
$\epsilon_{\downarrow}+U$ and  $\epsilon_{\uparrow}+U$ are 
outside the energy window opened by the applied voltage.

\begin{figure}[h]
\centerline{\includegraphics[width=6.8cm,angle=270]{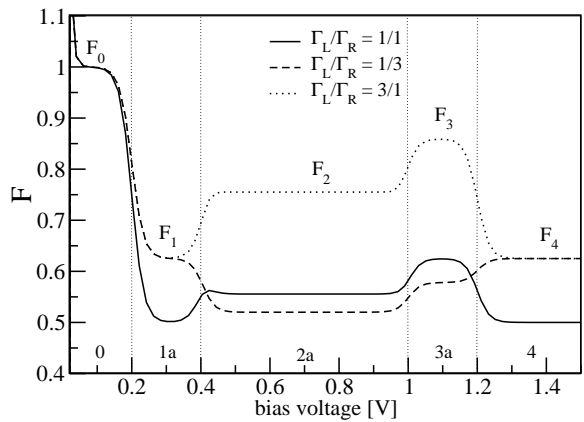}}
\caption{Fano factor vs. bias voltage for for the same parameters as in 
  Fig.~\ref{fig:i-s}, namely $T=100 {\rm K}$, 
  $\epsilon_{\downarrow}=0.1 {\rm eV}$,$\epsilon_{\uparrow}=0.2 {\rm eV}$,
  $U=0.4 {\rm eV}$, and $\Gamma_{\rm L} / \Gamma_{\rm R}=$ 1/1 (solid line), 1/3
  (dashed line), and 3/1 (dotted line). The labels $0,1a,2a,3a,4$ refer
to the cases listed in Table~\ref{tab:table-plateaus}}
\label{fig:fano}
\end{figure}

We now turn to the discussion of the Fano factor $F=S/2eI$.
In Fig.~\ref{fig:fano} we show the Fano factor as a function of bias voltage
for the same parameters as in Fig.~\ref{fig:i-s}.
Again we show three curves with $\Gamma_{\rm L}/\Gamma_{\rm R}=1/1$ 
(solid line), $1/3$ (dashed line) and $3/1$ (dotted line).
At small bias, $eV \ll k_BT$, the noise is dominated by thermal noise, 
described by the well-known hyperbolic cotangent behavior which leads
to a divergence of the Fano factor.\cite{blanter,loss}
The plateau for bias voltages below $0.2 {\rm V}$ corresponds to the 
Coulomb-blockade regime, where transport is exponentially suppressed
(case $0$ in Table~\ref{tab:table-plateaus}). 
In the region between $0.2 {\rm V}$ and $0.4 {\rm V}$ (case $1a$) transport 
through only one spin state ($\downarrow$) is possible.  
The Fano factor $F_1$ for this case has been derived earlier in 
Ref.~\onlinecite{nazarov}.
For very large bias (region $F_4$), all states of the molecular level are 
involved in transport, and the Fano factor $F_4$ is again identical to the
well-known~\cite{blanter} formula for transport through a resonant 
level in the absence of Coulomb charging energy.

For the regions in between, corresponding to the cases $2a$ and $3a$, the
Fano factor is different. The expression for $F_2$ has been very 
recently derived in Ref.~\onlinecite{bagrets}, while $F_3$,
corresponding to region $3a$, is, to the best of knowledge, a new result.

On similar grounds as for the current and the noise, we find that
the expressions for the Fano factors $F_2$ and $F_3$ are not invariant 
under exchange of $\Gamma_{\rm L}$ and $\Gamma_{\rm R}$ alone, and
also not invariant under reverse of the bias voltage alone.
This is clearly seen in Fig.~\ref{fig:fano} in  the very 
different plateau heights of the dotted and dashed curves,
corresponding to  exchange of $\Gamma_{\rm L}$ and $\Gamma_{\rm R}$.
Furthermore, we see that all plateau heights lie between $1/2$ and $1$, and 
that the Fano factor, in general, is a nonmonotonic function of the bias
voltage.
We find that $F_1 = F_4$ and $F_3 \geq F_2$ always hold, whereas 
$F_{1} \geq F_2$ for $\Gamma_{\rm L}/\Gamma_{\rm R} \leq 1/\sqrt{2}$
and $F_{1} \geq F_3$ for $\Gamma_{\rm L}/\Gamma_{\rm R} \leq 1/2$ only.

As a consequence, the pattern of the plateau sequence, in particular the 
relative height of the plateaus $F_2$ and $F_3$ compared to $F_1=F_4$, 
indicates not only the presence of an interaction or charging energy,
but also gives {\em overcomplete}
information on the ratio $\Gamma_{\rm L}/\Gamma_{\rm R}$
of the coupling strengths. This could be used in experiments to 
determine these parameters in a consistent way.  In an experiment
observing more than one plateau, the overcompleteness would give
narrow constraints (due to experimental uncertainty) on whether 
a single interacting  level can explain the measured values. Of
course, it is always possible to fit $n$ plateaus with $n$
noninteracting levels and different couplings per level, so an
absolute decision on the presence of interactions is not possible
without the additional application of a gate voltage.
However, if a fit with an interacting level is feasible, 
the principle of parsimony should favor the model with fewer parameters.

\begin{figure}[h]
\centerline{\includegraphics[width=6.8cm,angle=270]{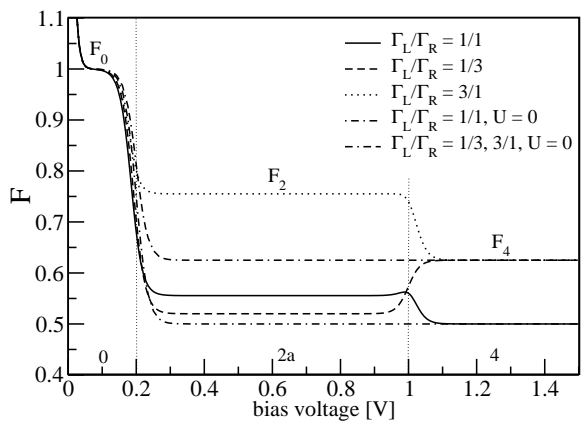}}
\caption{Fano factor vs. bias voltage in the absence of spin splitting.
  We chose $T=100 {\rm K}$, $\epsilon_{\downarrow}=\epsilon_{\uparrow}=0.1 
  {\rm eV}$, $U=0.4 {\rm eV}$, and $\Gamma_{\rm L} / \Gamma_{\rm R}=$ 1/1 
  (solid line), 1/3 (dashed line), and 3/1 (dotted line).
Also shown are the two cases for the corresponding noninteracting
system ($U=0$).}
\label{fig:fano-deg}
\end{figure}
In the absence of a spin splitting, $\epsilon_\uparrow=\epsilon_\downarrow$,
the number of plateaus is reduced.
An example for the Fano factor as a function of bias voltage is displayed in 
Fig.~\ref{fig:fano-deg}.
The solid curve is for symmetric, and the dotted and dashed 
curves are for asymmetric coupling.
Again, a nonmonotonic sequence of plateau heights, $F_4 > F_2$, indicates
the presence of Coulomb charging and 
$\Gamma_{\rm L}/\Gamma_{\rm R} < 1/\sqrt{2}$.
For comparison we remark that for both negligible spin splitting and zero 
charging energy ($U=0$), only the plateaus $F_0$ and $F_4$ appear, the 
plateau heights are invariant under reverse of bias, and no nonmonotonic 
behavior is seen, as also shown in Fig.~\ref{fig:fano-deg}.

Finally, we briefly comment on the steps connecting the plateaus of
the Fano factor. In the present first-order approximation the steps
are solely  broadened by temperature.
In addition, we observe that sometimes a peak in the Fano factor shows up.
This happens, for example, at the step between $F_1$ and $F_2$ for
$\Gamma_{\rm L} = \Gamma_{\rm R}$, as seen in Fig.~\ref{fig:fano}.
The peak height, $0.5625$, exceeds that of the adjacent plateaus, $0.5$ and 
$5/9$. These features can even appear in the regime of 
negligible Coulomb charging energy, as previously shown
in Ref.~\onlinecite{haug}. However, we point out that the
behavior at the steps is going to be strongly affected by second-order 
tunneling events, so-called cotunneling. For example,
the current steps will show additional broadening due to the
intrinsic linewidth $\Gamma$. 
On the other hand, the plateau values
in regimes 1 - 4 will not be affected by second-order effects.

In summary, we presented a theory of transport through a molecule 
or quantum dot that allows for a systematic perturbation expansion of current 
and shot noise in the coupling strength between molecular orbital
(dot level) and electrodes.
We analyzed first-order transport in detail.
In a complete overview, we derived analytic expressions for current, noise, and
the Fano factor for all possible situations involving a single
interacting level.
In particular, we discussed how the sequence of Fano factor plateaus can 
provide information about the asymmetry ratio 
of the coupling strengths as well as the presence of Coulomb charging energy.

{\em Acknowledgments.}
We enjoyed interesting and helpful discussions with A. Braggio, G. Johansson,
H. Schoeller, H. Weber, and A. Zaikin,
as well as financial support by the DFG via the Center for 
Functional Nanostructures and the Emmy-Noether program.

\end{document}